\documentclass[prl,twocolumn,amsmath,amssymb,showpacs]{revtex4}
\usepackage{graphicx}
\usepackage{dcolumn}
\usepackage{bm}

\begin{document}

\title{Nuclear spin dynamics in the quantum regime of a single-molecule magnet}
\author{A. Morello$^{1,*}$, O. N. Bakharev$^1$, H. B. Brom$^1$, R. Sessoli$^2$, and L. J. de Jongh$^1$}

 \affiliation{$^1$Kamerlingh Onnes Laboratory, Leiden University, P.O. Box 9504, 2300RA Leiden, The Netherlands.\\
 $^2$Department of Chemistry, University of Florence {\&} INSTM, via della
Lastruccia 3, 50019 Sesto Fiorentino (FI), Italy.}
\date{\today}

\begin{abstract}
We show that the nuclear spin dynamics in the single-molecule
magnet Mn$_{12}$-ac below 1 K is governed by quantum tunneling
fluctuations of the cluster spins, combined with intercluster
nuclear spin diffusion. We also obtain the first experimental
proof that - surprisingly - even deep in the quantum regime the
nuclear spins remain in good thermal contact with the lattice
phonons. We propose a simple model for how $T$-independent
tunneling fluctuations can relax the nuclear polarization to the
lattice, that may serve as a framework for more sophisticated
theories.
\end{abstract}


\maketitle

Single-molecule magnets (SMMs) are nanometer-sized high-spin
molecular clusters organized in a crystalline array, which sets
the direction for the anisotropy axis \cite{gatteschi94S}.
Reversal of the cluster spin can occur either classically, by
thermal activation, or quantum mechanically, by tunneling through
the barrier \cite{gatteschi94S,thomas96N}. SMMs are attractive
model systems to study the effects of coupling magnetic qubits to
the environment (nuclear moments, phonons), with the associated
problems of decoherence and the limits of quantum mechanics at the
large scale \cite{stamp04PRB,leggett02JPCM}. For both aspects the
hyperfine coupling between cluster spin and nearby nuclear spins
is expected to play a crucial but subtle role: since this coupling
is many orders larger than the quantum tunneling splitting, a
\textit{static} hyperfine interaction completely blocks tunneling.
Contrariwise, by considering it as a \textit{dynamic bias} that
sweeps the electron spin levels through the tunneling resonance,
Prokof'ev and Stamp (PS) have argued that this interaction in fact
\textit{promotes} incoherent tunneling events
\cite{prokof'ev96JLTP}.

Experimentally, although time-dependent magnetization experiments
\cite{wernsdorfer00PRL} showed a $\sqrt{t}$ dependence and isotope
effects agreeing with the PS predictions \cite{prokof'ev98PRL},
fundamental aspects of the spin-dynamics, like the essential role
of nuclear spin diffusion, remain to be verified. Further, in the
PS model the quantum relaxation of the cluster spin is to the
nuclear spin bath and is expected to be many orders of magnitude
faster than conventional spin-lattice relaxation to phonons. A
crucial test is thus whether or not the experimental nuclear
polarization relaxes to the lattice (phonon) temperature, even at
such low $T$ that only electron spin \textit{tunneling}
fluctuations are left (``quantum regime''). Interestingly, whereas
in order to relax to the lattice the nuclear spins generally need
electron spin fluctuations, in the quantum regime those same
nuclei would provide the only source for such fluctuations via the
PS nuclear-spin-mediated quantum tunneling model. It is by no
means obvious how such a $T$-independent process could establish
thermal equilibrium between spins and lattice. Here we report a
NMR study of the dynamics of $^{55}$Mn nuclei in
[Mn$_{12}$O$_{12}$(O$_{2}$CMe)$_{16}$(H$_{2}$O)$_{4}$]
(Mn$_{12}$-ac), which experimentally answers the above points and
poses a crucial test for a realistic description of the coupling
between a magnetic qubit and its environment.

Mn$_{12}$-ac is the SMM with the highest anisotropy barrier ($
\sim 65$ K) discovered so far; its core is composed of 4 Mn$^{4+}$
ions (electron spin $s = 3/2$), and 8 Mn$^{3+}$ ions ($s = 2$) in
two inequivalent crystallographic sites. The intracluster
superexchange interactions lead to a total spin $S = 10$ for the
cluster. Below $T \approx 3$ K the electron spins are effectively
frozen along the anisotropy axis, thereby enabling $^{55}$Mn NMR
even in zero applied field, by exploiting the local hyperfine
field $B_{\mathrm{hyp}}$ felt by the nuclei. This allows the use
of nuclear spins as local probes for the fluctuations of the
cluster spin by studying the nuclear spin-lattice relaxation
(NSLR) and the transverse spin-spin relaxation (TSSR),
\textit{without disturbing the zero-field tunneling resonance}. We
have chosen the resonance line of the $^{55}$Mn nuclei in Mn$^{4 +
}$ ions, having a central Larmor frequency $\omega_{N}/2\pi
\approx 230$ MHz and a relatively small quadrupolar splitting
\cite{kubo02PRB}. The experiments were performed on Mn$_{12}$-ac
crystallites, cast in Stycast 1266 epoxy and oriented in 9.4 T
magnetic field at room temperature. The NMR coil with the sample
was placed inside the elongated tail of the plastic mixing chamber
of a specially designed dilution refrigerator. This allows a
continuous flow of $^{3}$He around the sample and assures
excellent thermalization. The NMR signal was detected by spin-echo
technique, with typical duration $t_{\pi / 2}$ = 10 $\mu $s for
the 90$^{\circ}$ pulse. Since the $^{55}$Mn nuclei have spin $I$ =
5/2, the recovery of the nuclear magnetization, $M_z(t)$, after an
inversion pulse obeys \cite{suter98JPCM}: $M_z(t)/M_z(\infty ) = 1
- [(100/63)\exp(-30Wt) + (16/45)\exp(-12Wt) + (2/35)\exp(-2Wt)]$,
where $W$ is the NSLR rate \cite{T1vsW} [Fig. 1(c), solid lines].
The TSSR rate $T_{2}^{-1}$ is obtained by a single exponential fit
of the decay of transverse magnetization,
$M_{xy}(t)=M_{xy}(0)\exp(-t/T_{2})$, except at the lowest $T$
where also a gaussian component $T_{2G}^{-1}$ needs to be
included, yielding
$M_{xy}(t)=M_{xy}(0)\exp(-t/T_{2})\exp[-0.5(t/T_{2G})^2]$ [Fig.
1(b), solid lines].

\begin{figure}[t]
\includegraphics[width=8.5cm]{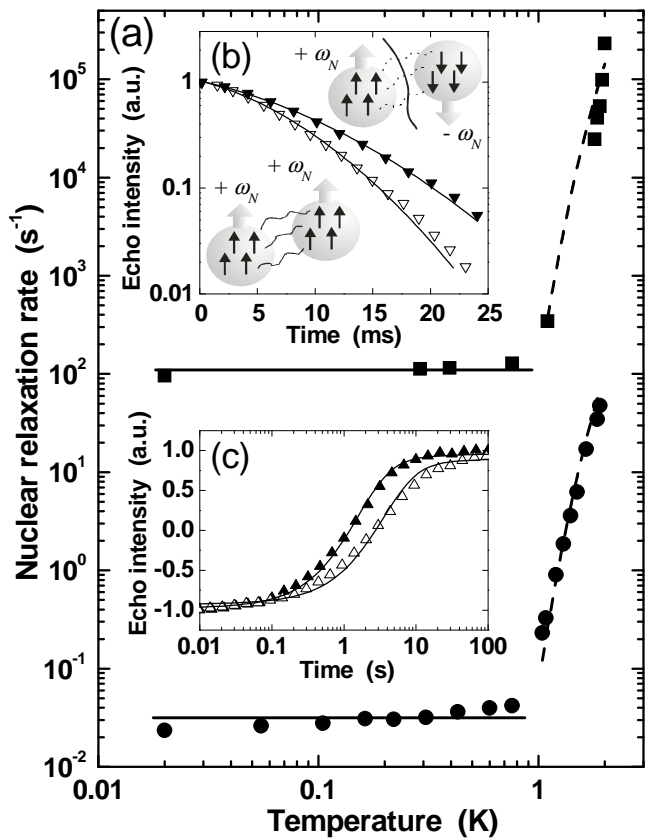}
\caption{(a) $T$-dependence of the NSLR ($\bullet$) and TSSR
($\blacksquare$) rates in zero field and ZFC sample. (b) Decay of
echo intensity at $T = 20$ mK in ZFC ($\blacktriangledown$) and FC
($\triangledown$) sample. Inset: in the ZFC sample half of the
nuclear spins (black arrows) has Larmor frequency $-\omega_{N}$
instead of $+\omega_{N}$ because of the reversed orientation of
the cluster spin (gray). (c) Recovery of nuclear magnetization
after an inversion pulse, at $T = 20$ mK in ZFC ($\blacktriangle$)
and FC ($\vartriangle$) sample.}
\end{figure}

Between 1 and 2 K, both the NSLR and the TSSR show a roughly
exponential $T$-dependence [dashed curves in Fig. 1(a)], which is
well understood in terms of the fluctuations of $B_{\mathrm{hyp}}$
produced by thermal activation of the electron spin levels
\cite{furukawa01PRB,goto03PRB,morello03POLY}. The NSLR rate can be
obtained from the spectral density at $\omega = \omega_N$ of the
transverse component of the fluctuating part of
$B_{\mathrm{hyp}}$, with the implicit assumption that
$B_{\mathrm{hyp}}$ fluctuates around its average direction (which
coincides with the molecule's anisotropy axis) but does not flip
over, as in a tunneling event. Extrapolating the observed high-$T$
NSLR to the mK range would lead to astronomically long relaxation
times. In a preliminary work \cite{morello03POLY} we observed
that, upon cooling down to 20 mK, the NSLR saturates to a roughly
$T$-independent plateau, indicating that only fluctuations due to
quantum tunneling within the ground doublet are contributing to
the relaxation. The crossover between the thermally activated
regime and the quantum regime [Fig. 1(a), solid lines] is clearly
visible at 0.8 K, in agreement with magnetization experiments
\cite{chiorescu00PRL}.

The value $W_{0} \approx 0.03$ s$^{-1}$ of the NSLR found below
0.8 K is surprisingly high, considering that the relaxation of the
global magnetization in Mn$_{12}$-ac takes years at low $T$. On
the other hand, it is well known that any real sample of
Mn$_{12}$-ac contains a fraction of fast-relaxing molecules
(FRMs), which are characterized by one or two distorted local
anisotropy axes for the Mn$^{3+}$ ions \cite{sun99CC}; for those
molecules the barrier is reduced to 35 K or even 15 K
\cite{wernsdorferU}, yielding much faster tunneling dynamics. At
the same time, however, we have verified that the observed NMR
signal comes from nuclei in standard, slow-relaxing molecules,
even though the electron spin of such molecules remains frozen
during the experiment. The fluctuating dipolar field produced by a
tunneling FRM on the nuclei of neighboring (frozen) molecules is
far too small to account for the observed NSLR, so we have
suggested \cite{morello03POLY} that the relaxation mechanism
should involve \textit{intercluster} nuclear spin diffusion (not
included in Refs. \cite{prokof'ev96JLTP,prokof'ev98PRL}), linking
nuclei in frozen molecules to those in FRMs. By studying the
magnetization dependence of $T_{2}^{ - 1}$ we can now provide
strong evidence for the proposed mechanism. When comparing the
TSSR in a demagnetized, zero-field cooled (ZFC) sample (where the
cluster spins are randomly oriented up or down) with a saturated,
field-cooled (FC) sample (where all spins have the same
direction), we find that the FC sample has a faster TSSR, with a
ratio $T_{2G}^{-1}\mathrm{(FC)} / T_{2G}^{-1}\mathrm{(ZFC)}
\approx 1.35$, very close to $\sqrt{2}$ [Fig. 1(b)]. In terms of
intercluster spin diffusion this has a simple explanation: in a FC
sample the nuclei in equivalent crystallographic sites of
different molecules have the same Larmor frequency, $\omega_{N}$,
thus flip-flop transitions are possible with all neighbors. In a
ZFC sample the nuclei are divided in two groups having Larmor
frequencies $+\omega_{N}$ or $-\omega_{N}$, depending on the local
spin orientation; for the nuclear dipole-dipole interaction this
is equivalent to having diluted the FC system by a factor 2,
yielding a $\sqrt{2}$ times smaller TSSR \cite{sqrt2} [see inset
of Fig. 1(b)]. The presence of a predominantly gaussian component
as found in the TSSR at low $T$ confirms the importance of nuclear
dipolar couplings.

Further insight in the relationship between the dynamics of the
central quantum spin and the nuclei is provided by the field
dependence of the NSLR. Applying an external field $B_{z}$
parallel to the anisotropy axis destroys the resonance condition
for tunneling, thereby hindering the fluctuations needed for the
NSLR; this explains the pronounced peak in $W(B_{z})$ found around
zero field as shown in Fig. 2(a). In comparing ZFC and FC sample,
it is seen that both the width of the resonance and the zero-field
value are quite different. In particular, one may conclude that
there are more tunneling events at zero field in the ZFC sample,
as could be seen already from the difference in nuclear inversion
recovery [Fig. 1(c)]. Such an observation, which is obviously
impossible to obtain by means of ``macroscopic'' magnetization
measurements, should provide a critical test for more detailed
models of the NSLR. The signature of tunneling fluctuations at the
first level crossing around $B_{z} \approx 0.5$ T, i.e. when the
spin states with $S_{z} = +10, +9, \ldots$ come in resonance with
$S_{z} = -9, -8, \ldots$ , becomes visible as a small peak in
$W(B_{z})$ only upon warming up to $T = 0.72$ K, i.e. close to the
border with the thermally activated regime [Fig. 2(b)] \cite{B1C}.

\begin{figure}[t]
\includegraphics[width=8.5cm]{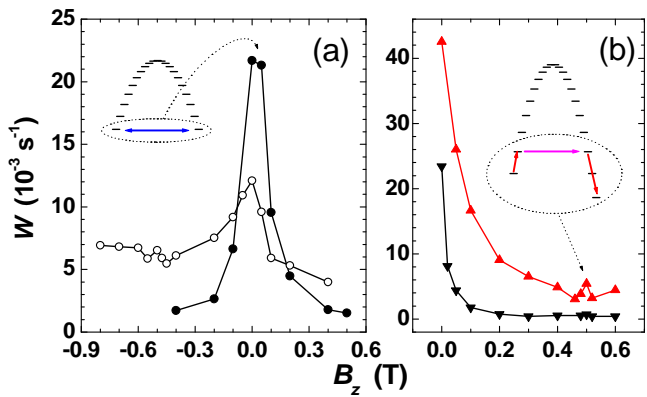}
\caption{(Color online) (a) Longitudinal field dependence of the
NSLR rate $W(B_{z})$ at $T = 20$ mK in the ZFC ($\bullet$) and FC
($\circ$) sample. The measuring frequency was set to
$\omega_{N}(B_{z}) / 2\pi = 230 + 10.57 B_{z}$ MHz. (b) $W(B_{z})$
in ZFC sample at $T$ = 20 mK ($\blacktriangledown$) and $T$ = 720
mK ($\blacktriangle$): a small peak is visible at the first level
crossing $B_{z}\approx 0.5$ T only at the highest temperature. In
this dataset we used $\omega_{N}(B_{z}) / 2\pi = 231 + 10.57
B_{z}$ MHz, which better matches the center of the NMR line at
high $T$. The insets show a sketch of the electronic level scheme
with the observed transitions.}
\end{figure}

Finally, we address another essential aspect of the dynamics of
the coupled system of nuclear and cluster spins, so far not
studied theoretically or experimentally, namely: ``what is the
nuclear spin temperature''? In other words, is the nuclear spin
polarization indeed relaxing to an equilibrium value dictated by
the lattice phonons, which are in thermal contact with the $^3$He
bath at temperature $T_{\mathrm{bath}}$? In that case the
intensity of the NMR signal as a function of temperature should
obey the Curie law $M_z(T)=K /T$. The calibration factor $K$ can
be defined at the highest $T$ by assuming that there the nuclear
spin temperature $T_{\mathrm{nucl}}$ equals $T_{\mathrm{bath}}$,
and then be used to convert the NMR signal intensity into an
equivalent $T_{\mathrm{nucl}}$ while cooling down the system. As
shown in Fig. 3(a), we find that $T_{\mathrm{nucl}}$ indeed
follows the time evolution of $T_{\mathrm{bath}}$, the small
discrepancy below 0.2 K being most probably due to heating effects
of the NMR pulses. Data taken with a lower pulse rate [Fig. 3(b)]
demonstrate that, even below $0.1$ K, the nuclei always closely
follow the evolution of $T_{\mathrm{bath}}$. This direct
experimental proof of an energetic contact between nuclear spins
and phonons confirms earlier results from the (field-dependent)
low-$T$ specific heat \cite{mettes01PRB}, in which sizable amounts
of nuclear and electron magnetic entropy were observed to be
removed below 0.5 K. Since there is no relevant direct energetic
coupling between phonons and nuclei, the thermalization of the
nuclear spin system must involve the interplay with the electron
spins and their coupling to the lattice, even in the
$T$-independent quantum tunneling regime.

\begin{figure}[t]
\includegraphics[width=8.5cm]{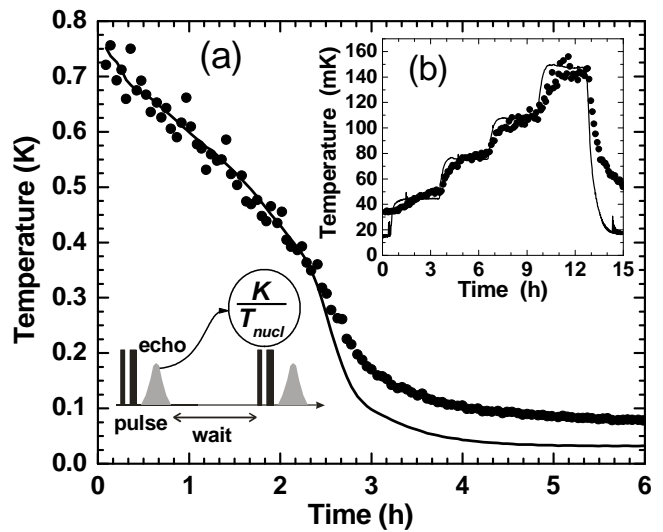}
\caption{Comparison of bath temperature $T_{\mathrm{bath}}$ (solid
lines) and nuclear spin temperature $T_{\mathrm{nucl}}$ (circles),
while cooling down (a) and while applying step-like heat loads
(b). The waiting time between NMR pulses (see inset) was 60 s in
(a) and 180 s in (b). Both datasets are at zero field in ZFC
sample.}
\end{figure}

A basic question to answer is what happens to those nuclei that
belong to a molecule where a tunneling event takes place (in our
case a FRM), assuming that the neighboring molecules are frozen.
For the ease of discussion, we shall consider $N$ nuclear spins $I
= 1/2$ per cluster, subject to a hyperfine field
$B_{\mathrm{hyp}}$ parallel to the anisotropy axis of the
molecule: the latter assumption simulates the real situation for
$^{55}$Mn in Mn$_{12}$-ac. The standard way of calculating the
rate of transition between nuclear Zeeman levels as a consequence
of a perturbing fluctuating field is useless here, since the
Zeeman levels themselves completely change after each electron
spin flip, so perturbation theory is not applicable. A more
realistic approach is to recall that each electron spin level is
split by hyperfine interactions into a quasi-continuum manifold of
levels \cite{kubo02PRB,hartmann96IJMPB} that can be labeled by the
local nuclear polarization $\Delta N = N^{ \uparrow }-N^{
\downarrow }$, which yields a hyperfine bias $\xi_{N}$ (typically
$\sim 0.1$ K). Since the hyperfine fields before and after the
flip of the cluster spin are just antiparallel, the manifolds of
Zeeman levels on either side of the anisotropy barrier are simply
the mirror of each other. Moreover, since the tunneling traversal
time is much shorter than $1/\omega_{N}$, the probability that a
nuclear spin would coflip with the electron spin is negligible.
This implies that the only relevant tunneling transitions are
those that do not require any nuclear coflip, thus $\Delta N$ =
const. \cite{prokof'ev96JLTP}. Considering the small additional
bias $\xi_{D}$ due to dipolar fields from neighboring cluster
spins, the tunneling transition with $\Delta N$ = const. requires
an initial hyperfine bias such that $\xi_{N}=\xi_{D}$. Once the
molecule has tunneled, the hyperfine bias becomes $\xi_{N} =
-\xi_{D}$ since the nuclear polarization is unchanged but
$B_{\mathrm{hyp}}$ is reversed; the new local hyperfine energy can
then be redistributed to other nuclei via intercluster spin
diffusion until the equilibrium within the nuclear spin bath is
achieved. In this way the effect of tunneling is the
``conversion'' of dipolar into hyperfine energy and vice versa.

Our data show that this description is still insufficient: to
obtain a nuclear magnetization in thermal equilibrium with the
lattice, tunneling events must be accompanied by
creation/annihilation of phonons. In our opinion a crucial role
may be played by the Waller mechanism \cite{waller32ZP}, i.e. the
change in the dipolar field when the distance between neighboring
molecules is modulated by lattice vibrations. Even at very low $T$
we can expect the existence of low-energy phonon modes that
correspond to displacements of the clusters with respect to each
other. Here we consider the cluster cores as rigid objects within
the soft matrix (with Debye temperature $\theta_{D}\approx 20$ K)
of the ligand molecules, an approach successfully used to account
for the M\"{o}ssbauer recoil-free fractions of metal cluster
molecules \cite{paulus01PRB}. The modulation of the dipolar field,
whereby the total bias may sweep back and forth through the
tunneling resonance, can thus provide a probability of incoherent
tunneling with emission or absorption of phonons, whose energy
would be released or extracted from the nuclear spins in the way
described above, i.e. using the tunneling of electron spins as
intermediary. The detailed balance between emission and absorption
may then provide the equilibration of nuclear spin and lattice
temperatures.

In conclusion, we have shown that the nuclear spin dynamics in
Mn$_{12}$-ac below 0.8 K is driven by tunneling fluctuations of
the cluster electron spin, in combination with intercluster
nuclear spin diffusion and thermal equilibrium between nuclear
spins and phonon bath; the latter aspect calls for the extension
of existing theories of incoherent quantum tunneling within the
ground doublet to include inelastic processes.

\begin{acknowledgements}
We acknowledge stimulating discussions with P. C. E. Stamp, B. V.
Fine, I. S. Tupitsyn, F. Luis, F. Borsa and Y. Furukawa. We also
thank W. Wernsdorfer for providing us unpublished data. This work
is part of the research program of the "Stichting FOM" and is
partially financed by the European Community under contract no.
MRTN-CT-2003-504880 "QUEMOLNA".
\end{acknowledgements}

$^*$ Present address: Department of Physics and Astronomy,
University of British Columbia, Vancouver B.C. V6T 1Z1, Canada.
E-mail: morello@physics.ubc.ca


\begin{thebibliography}{10}

\bibitem{gatteschi94S}
{D. Gatteschi, A. Caneschi, L. Pardi, and R. Sessoli, Science
\textbf{265}, 1054 (1994); D. Gatteschi and R. Sessoli, Angew.
Chem. Int. Ed. \textbf{42}, 268 (2003).}

\bibitem{thomas96N}
{J. R. Friedman, M. P. Sarachik, J. Tejada, and R. Ziolo, Phys.
Rev. Lett.
  \textbf{76}, 3830 (1996); J. M. Hern\'{a}ndez, X. X. Zhang, F. Luis, J. Bartolom\'{e}, J. Tejada, and R. Ziolo, Europhys. Lett. \textbf{35}, 301
  (1996); L. Thomas, F. Lionti, R. Ballou, D. Gatteschi, R. Sessoli, and B. Barbara, Nature (London) \textbf{383}, 145 (1996).}

\bibitem{stamp04PRB}
{P. C. E. Stamp and I. S. Tupitsyn, Phys. Rev. B \textbf{69},
014401 (2004).}

\bibitem{leggett02JPCM}
{A. J. Leggett, J. Phys.: Condens. Matter \textbf{14}, R415
(2002).}

\bibitem{prokof'ev96JLTP}
{N. V. Prokof'ev and P. C. E. Stamp, J. Low Temp. Phys.
\textbf{104}, 143
  (1996).}

\bibitem{prokof'ev98PRL}
{N. V. Prokof'ev and P. C. E. Stamp, Phys. Rev. Lett. \textbf{80},
5794
  (1998).}

\bibitem{wernsdorfer00PRL}
{W. Wernsdorfer, A. Caneschi, R. Sessoli, D. Gatteschi, A. Cornia,
V. Villar, and C. Paulsen, Phys. Rev. Lett. \textbf{84}, 2965
(2000).}

\bibitem{kubo02PRB}
{T. Kubo, T. Goto, T. Koshiba, K. Takeda, and K. Awaga, Phys. Rev.
B \textbf{65}, 224425 (2002).}

\bibitem{suter98JPCM}
{A. Suter, M. Mali, J. Roos and D. Brinkmann, J. Phys.: Condens.
Matter
  \textbf{10}, 5977 (1998).}

\bibitem{T1vsW}
{In the simple case of a spin 1/2, where the NSLR is described by
a single
  exponential, $W$ is related to the relaxation time $T_1$ by $2W=T_1^{-1}$.}

\bibitem{furukawa01PRB}
{Y. Furukawa, K. Watanabe, K. Kumagai, F. Borsa, and D. Gatteschi,
Phys. Rev. B
  \textbf{64}, 104401 (2001).}

\bibitem{goto03PRB}
{T. Goto, T. Koshiba, T. Kubo, and K. Awaga, Phys. Rev. B
\textbf{67}, 104408
  (2003).}

\bibitem{morello03POLY}
{A. Morello, O. N. Bakharev, H. B. Brom, and L. J. de Jongh,
Polyhedron
  \textbf{22}, 1745 (2003).}

\bibitem{chiorescu00PRL}
{L. Bokacheva, A. D. Kent, and M. A. Walters, Phys. Rev. Lett
\textbf{85}, 4803
  (2000); I. Chiorescu, R. Giraud, A. G. M. Jansen, A. Caneschi and B. Barbara,
  \textit{ibid}. 4807.}

\bibitem{sun99CC}
{Z. Sun, D. Ruiz, N. R. Dilley, M. Soler, J. Ribas, K. Folting, M.
B. Maple, G. Christou, and D. N. Hendrickson, Chem. Comm. 1973
(1999).}

\bibitem{wernsdorferU}
{W. Wernsdordfer (unpublished).}

\bibitem{sqrt2}
{In reality also in a ZFC sample some agglomerates of
  clusters may locally have the same orientation, whereby this ratio becomes $<\sqrt{2}$, as observed.}

\bibitem{B1C}
{Notice that $B_z \approx 0.5 $ T is the level crossing field only
for the
  majority of molecules with 65 K barrier, whereas the crossing fields for FRMs
  are distributed between 0.2 and 0.4 T and do not give sharply visible
  features.}

\bibitem{mettes01PRB}
{F. Luis, F. L. Mettes, J. Tejada, D. Gatteschi, and L. J. de
Jongh, Phys. Rev. Lett. \textbf{85}, 4377 (2000); F. L. Mettes, F.
Luis, and L. J. de Jongh, Phys. Rev. B \textbf{64}, 174411 (2001);
M. Evangelisti, F. Luis, F. L. Mettes, N. Aliaga, G. Arom\'{i}, J.
J. Alonso, G. Christou, and L. J. de Jongh, Phys. Rev. Lett.
\textbf{93}, 117202 (2004).
  (2001).}

\bibitem{hartmann96IJMPB}
{F. Hartmann-Boutron, P. Politi, and J. Villain, Int. J. Mod.
Phys. B
  \textbf{10}, 2577 (1996).}

\bibitem{waller32ZP}
{I. Waller, Z. Phys. \textbf{79}, 370 (1932).}

\bibitem{paulus01PRB}
{P. M. Paulus, A. Goossens, R. C. Thiel, A. M. van der Kraan, G.
Schmid, and L. J. de Jongh, Phys. Rev. B \textbf{64}, 205418
(2001).}

\end{thebibliography}
\end{document}